\documentclass{aa}
\usepackage{graphicx}
\usepackage{txfonts}
\newcommand  \kms      {\ifmmode {\rm km\,s}^{-1} \else km\,s$^{-1}$\fi}

\newcommand  \cmii     {\hbox{cm$^{-2}$}}
\newcommand  \ergs     {\ifmmode {\rm ergs\,s}^{-1} \else ergs s$^{-1}$\fi}
\newcommand  \ergcms   {\ifmmode {\rm ergs\,cm}^{-2}\,{\rm s}^{-1}
                        \else ergs\,cm$^{-2}$\,s$^{-1}$\fi}
\newcommand  \ergcmsA {\ifmmode{\rm ergs\,cm}^{-2}\,{\rm s}^{-1}\,{\rm\AA}^{-1}
                        \else ergs\,cm$^{-2}$\,s$^{-1}$\,\AA$^{-1}$\fi}
\newcommand \ergcmsHz {\ifmmode{\rm ergs\,cm}^{-2}\,{\rm s}^{-1}\,{\rm Hz}^{-1}
                        \else ergs\,cm$^{-2}$\,s$^{-1}$\,Hz$^{-1}$\fi}
\newcommand  \phcms    {\ifmmode {\rm ph\,cm}^{-2}\,{\rm s}^{-1}
                        \else ,ph\,cm$^{-2}$\,s$^{-1}$\fi}
\newcommand  \phcmsA   {\ifmmode {\rm ph\,cm}^{-2}\,{\rm s}^{-1}\,{\rm\AA}^{-1}
                        \else ph\,cm$^{-2}$\,s$^{-1}$\,\AA$^{-1}$\fi}
\def\micron{\ifmmode \mu{\rm m} \else $\mu$m\fi}
\def\kms{\ifmmode {\rm km\,s}^{-1} \else km\,s$^{-1}$\fi}
\def\Hubble{\ifmmode {\rm km\,s}^{-1}\,{\rm Mpc}^{-1}
        \else km\,s$^{-1}$\,Mpc$^{-1}$\fi}
\def\ergsec{\ifmmode {\rm ergs\;s}^{-1} \else ergs s$^{-1}$\fi}
\def\ergscm{\ifmmode {\rm ergs\,s}^{-1}\,{\rm cm}^{-2}
          \else ergs\,s$^{-1}$\,cm$^{-2}$\fi}
\def\ergscmA{\ifmmode {\rm ergs\,s}^{-1}\,{\rm cm}^{-2}\,{\rm \AA}^{-1}
          \else ergs\,s$^{-1}$\,cm$^{-2}$\,\AA$^{-1}$\fi}
\def\ergscmHz{\ifmmode {\rm ergs\,s}^{-1}\,{\rm cm}^{-2}\,{\rm Hz}^{-1}
          \else ergs\,s$^{-1}$\,cm$^{-2}$\,Hz$^{-1}$\fi}
\def\Msun{\ifmmode M_{\odot} \else $M_{\odot}$\fi}
\def\Lsun{\ifmmode L_{\odot} \else $L_{\odot}$\fi}
\def\qo{\ifmmode q_{0} \else $q_{0}$\fi}
\def\Ho{\ifmmode H_{0} \else $H_{0}$\fi}
\def\ho{\ifmmode h_{0} \else $h_{0}$\fi}
\def\qo{\ifmmode q_{0} \else $q_{0}$\fi}
\def\ao{\ifmmode a_{0} \else $a_{0}$\fi}
\def\to{\ifmmode t_{0} \else $t_{0}$\fi}
\def\Halpha{\ifmmode {\rm H}\alpha \else H$\alpha$\fi}
\def\Hbeta{\ifmmode {\rm H}\beta \else H$\beta$\fi}
\def\hb{\ifmmode {\rm H}\beta \else H$\beta$\fi}
\def\Hgamma{\ifmmode {\rm H}\gamma \else H$\gamma$\fi}
\def\Hdelta{\ifmmode {\rm H}\delta \else H$\delta$\fi}
\def\Lya{\ifmmode {\rm Ly}\alpha \else Ly$\alpha$\fi}
\def\Lyb{\ifmmode {\rm Ly}\beta \else Ly$\beta$\fi}
\def\hi{\ifmmode \mbox{{\rm H}\,{\sc i}} \else H\,{\sc i}\fi}

\def\heii{He\,{\sc ii}\,$\lambda1640$}

\def\ciii{\ifmmode {\rm C}\,{\sc iii} \else C\,{\sc iii}\fi}
\def\civ{C\,{\sc iv}\,$\lambda1549$}

\def\oii{[O\,{\sc ii}]\,$\lambda3727$}
\def\oiii{[O\,{\sc iii}]\,$\lambda5007$}

\def\neiii{Ne\,{\sc iii}\,$\lambda3870$}

\def\o5007{[O\,{\sc iii}]\,$\lambda5007$}

\def \LOLX { ${\rm L_{[OIII]}}$/${\rm L_{2-10}}$ }
\def \LOIILX {${\rm L_{[OII]}}$/${\rm L_{2-10}}$}
\def  \kms         {\hbox{km s$^{-1}$}}          
\def  \ergs        {\hbox{erg s$^{-1}$}}              

\def  \cmii        {\hbox{cm$^{-2}$}}

\def  \La          {\ifmmode {\rm Ly}\alpha \else Ly$\alpha$\fi}
\def  \Ka          {\ifmmode {\rm K}\alpha \else K$\alpha$\fi}
\def  \Lb          {\ifmmode {\rm L}\beta \else L$\beta$\fi}
\def  \Ha          {\ifmmode {\rm H}\alpha \else H$\alpha$\fi}
\def  \Hb          {\ifmmode {\rm H}\beta \else H$\beta$\fi}
\def  \Pa          {\ifmmode {\rm P}\alpha \else P$\alpha$\fi}
\def  \CIIIb       {\ifmmode {\rm C}\,{\sc iii]}\,\lambda1909
                     \else C\,{\sc iii]}\,$\lambda1909$\fi}
\def  \CIV         {\ifmmode {\rm C}\,{\sc iv}\,\lambda1549
                     \else C\,{\sc iv}\,$\lambda1549$\fi}
\def  \MgII         {\ifmmode {\rm Mg}\,{\sc ii}\,\lambda2798
                     \else Mg\,{\sc ii}\,$\lambda2798$\fi}
\def  \OVI         {\ifmmode {\rm O}\,{\sc vi}\,\lambda1035
                     \else O\,{\sc vi}\,$\lambda1035$\fi}
\def \chandra  {{\it Chandra}}
\def \xmm      {{\it XMM-Newton}}

\begin{document}

\title{The Correlation of Narrow Line Emission and X-ray Luminosity in Active Galactic Nuclei }

\author{
Hagai Netzer\inst{1,2}
\and
V. Mainieri\inst{2}
\and
P. Rosati\inst{3}
\and
Benny Trakhtenbrot\inst{1}
}


\institute
                {School of Physics and Astronomy and the Wise
                Observatory, The Raymond and Beverly Sackler Faculty of
                Exact Sciences, Tel-Aviv University, Tel-Aviv 69978,
                Israel\\
                \email{netzer@wise.tau.ac.il}
  \and
Max-Planck-Institut f\"ur extraterrestrische Physik, Postfach 1312, 85741 Garching, Germany \\
\and
      European Southern Observatory, Karl-Schwarzschild-Strasse 2, 85748 Garching bei Muenchen, Germany
}

\date{Accepted, March 15 2006}

  \abstract
{}
{We combine emission line and X-ray luminosities for 45 sources 
from the \chandra\ Deep Field South (CDF-S),  and seven 
HELLAS sources, to obtain a new sample of 52 X-ray selected 
type-II active galactic nuclei (AGNs). Eighteen of our sources are very luminous with a typical, 
absorption-corrected 2--10
keV luminosity of $few \times 10^{44}$ \ergsec\ (type-II QSOs).
}
{
We compare the emission line properties of the new sources
with emission line and X-ray luminosities of known low redshift, mostly lower luminosity  AGNs
by using a composite spectrum.
}
{
 We find that
\LOLX\ and \LOIILX\ decrease with L(2-10 keV) such that
\LOLX$\propto {\rm L_{2-10}}^{-0.42}$.
The trend was already evident, yet neglected in past low redshift samples.
This lead to erroneous calibration of the line-to-X-ray luminosity in earlier  AGN samples.
The analysis of several type-I samples shows the same trend with a similar slope
but a median \LOLX\ which is larger by a factor of about two compared with optically selected
type-II samples. We interpret this shift as due to additional reddening
in type-II sources and comment in general on the very large extinction in many type-II objects
and the significantly smaller average reddening of the SDSS type-II AGNs.
The decrease of \LOLX\ with L(2--10 keV) is
large enough to suggest that a significant fraction of  high luminosity
high redshift type-II AGNs have very weak emission lines that may have
escaped detection in large samples. A related decrease of EW(\oiii) with optical continuum luminosity
is  demonstrated by an analysis of 12,000 type-I SDSS AGNs. 
The new correlations found here are important for deriving accurate luminosity functions 
for AGNs and their neglect may explain past discrepancies between emission line and
X-ray samples.
}
{}


 \maketitle

\section{Introduction}
The study of the space distribution and the luminosity function (LF) 
of active galactic nuclei (AGNs) has been the focus of much attention in recent
years. In particular, deep X-ray surveys (Hasinger, 2004; Ueda et al. 2003
and references therein) have been combined with large ground-based data sets, like the two degree 
field (2dF) galaxy redshift survey and
the Sloan Digital Sky Survey (SDSS) to study the LFs
of type-I and type-II AGNs. Such studies are essential for understanding AGN and galaxy evolution and to
explain the contribution of the various sub-groups to the cosmological X-ray background (CXB). 

Several  extensive studies of the differences between type-I and type-II sources
make use of the \oiii\ line luminosity. This line originates in the narrow line
region (NLR) which is thought to be of much larger dimensions  than the putative central torus.
Thus, the line emission is considered to be isotropic and to have a similar
luminosity distribution in type-I and type-II sources.
Work by Mulchaey et al. (1994; hereafter M94) and Alonso-Herrero, Ward and Kotilainen (1997; hereafter A97)
suggested a simple calibration scheme between the hard (2--10 keV) X-ray luminosity and L(\oiii). These papers,
and several others since, studied 
various correlations between L(\oiii), L(2--10 keV) and L(IR) in small samples of mostly
low luminosity AGNs.  According to these papers, the mean and the distribution
of L(\oiii)/L(2--10 keV) (hereafter \LOLX) in the two AGN groups is indistinguishable confirming
 the line isotropy assumption and suggesting
that the mean line extinction is also independent of orientation. 
This M94 scaling relationship was used as a standard optical-to-X-ray conversion tool
 in numerous other papers, including very recent ones (e.g. Vignali et al. 2004).

More recent studies made use of the large
SDSS sample to study the L(\oiii) and the line equivalent width (EW) distribution.
Some of the most extensive studies of this type  that are relevant to the present work are Zakamska et al. 
(2003; 2004), Simpson (2005; hereafter S05), Hao et al. (2005) and Heckman et al. (2005). 
 The Zakamska et al. papers made use of the M94 and A97 results and
demonstrated the similar distribution of L(\oiii) in the two AGN subgroups in the redshift
range $0.3<z<0.8$. 
They also show an almost perfect linear relationship between L(\oii) and L(\oiii),
thus the \oii\ line is also expected to show a luminosity independent ratio with L(2--10 keV),
L(\oii)/L(2--10 keV) (hereafter \LOIILX), similar to  \LOLX.
Heckman et al. (2005) further tested the M94 assumption and noted the change in 
the mean \LOLX\ between X-ray selected and emission line selected samples. The paper 
also notes the larger
range in this property in type-II sources due to X-ray obscuration.

The recent works by S05 and by Hao et al. (2005) focus on line LFs obtained from the SDSS
data set. S05 derived \oiii\ LFs  for the two AGN sub-groups and used the M94 scaling to compare them
with large X-ray samples. He further makes a detailed comparison with
the ``receding torus'' model (see e.g. Grimes et al. 2004; S05).
 The conclusion is that the model in its simplest form cannot 
explain the different LFs derived for the two groups.
Hao et al. (2005) focus on the similarity of the \Ha\ and \oiii\ LFs in two large samples of AGNs with
$0<z<0.15$.
 
Much of the uncertainty in deriving the L(\oiii) LF for type-II sources is the lack of reliable
measurements of this line in high redshift high luminosity sources. There are a handful of notable exceptions
(see Stern et al. 2002; Norman et al. 2002) but, so far, the numbers were far from enough to make a detailed study
of the population. Almost the only way to cure this deficiency is to combine X-ray fluxes and follow-up spectroscopy
in X-ray selected samples.
In this paper we adopt this approach and discuss the emission line spectrum of 52
  high redshift type-II AGNs, some with very high X-ray luminosity.
 Most of the sources were discovered in
 the 1Ms {\it Chandra} observation of the
Chandra Deep Field South (CDF-S; Giacconi et al. 2002; Rosati et
al. 2002) and seven others are obtained from the HELLAS2XMM survey (Fiore et al. 2003).
The new sample allows us to address the issue of the observed
vs. the  expected \LOLX\ and \LOIILX\ in type-I and type-II AGNs and to draw conclusions about the LF
of type-II sources at high redshifts.
\S2 describes the  observations and data sources 
used in this work and the procedure used for obtaining an ``effective L(\oiii)''
for our high redshift sample. We then show various new diagrams involving
L(\oiii), L(\oii) and L(2--10 keV)  and argue for  new
luminosity dependent correlations of narrow emission lines in type-II sources.
Finally in \S3 we discuss the new results, compare them with SDSS measurements of \oiii, and
address some of the consequences to AGN  LFs.

\section{X-ray and optical spectroscopy of new type-II AGNs}

\subsection{The new type-II AGN sample}
The new sample described in this paper is the result of a
spectroscopic follow-up of the 
CDF-S.  The spectroscopy and the
selection criteria used to identify type-I and type-II AGNs, are
explained in Szokoly et al. (2004). In short, we have obtained
spectroscopic redshifts for 168 X-ray sources of which 137 have both
reliable optical identification and redshift estimates. Out of them we
identified 52 type-II AGNs based on their X-ray obscuring column and and luminosity (Tozzi et al. 2006).
To be fully consistent with the type-II definition, we only consider sources
with N$_{\rm H} > 10^{22}$ cm$^{-2}$ and L(2--10 keV) $>10^{42}$ 
(the name ``type-II QSOs'' is reserved to
those sources with L(2--10 keV) $>10^{44}$ erg s$^{-1}$)\footnote{An additional source, XID$=62$ from Giacconi et
al. (2002), which would fit these selection criteria has been excluded
since the optical spectroscopy reveals its BAL-QSO nature}.
Two of those (CDF-S\,202 and CDF-S\,263) have already been discussed in previous papers
(Norman et al. 2002; Mainieri et al. 2005). 

Table 1 summarizes all relevant spectroscopic data for the type-II sources belonging to this 
sample (for more spectroscopic information see Szokoly et al. 2004).
Columns 1 and 2 give the object name and redshift and columns 6--14 give observed line fluxes for 
\Lya, \civ, \heii, \CIIIb, \MgII, \oii, \neiii, \hb\ and \oiii\footnote{Emission line fluxes
for 7 of the sources are also shown in a paper by Nagao, Marconi \& Maiolino that
was submitted to publication after the submission of our paper - see astroph-0508652}.
 Columns 3--5 provide information
on the X-ray spectrum: the photon spectral index $\Gamma$, the obscuring neutral column  N$_{\rm H}$ obtained from
the fit and the intrinsic (corrected for absorption) 2--10 keV luminosity. These
values have been obtained by a detailed spectral fitting procedure
(Tozzi et al. 2006) in which the default spectral model was a power
law with slope $\Gamma$, intrinsic redshifted absorber with a neutral hydrogen column of N$_{\rm H}$ \cmii,
 fixed Galactic absorption and an unresolved Fe emission line. We
also allowed for the presence of a scattered component at soft
energies with the same slope of the main power-law, and for a pure
reflection typical of Compton-thick AGN.
We have decided to omit five Compton thick sources defined here
as objects with $N_H > 10^{24}$ \cmii. These are likely to give highly uncertain,
probably erroneous intrinsic \LOLX.

\tiny
\begin{table}
\tiny
\caption{Emission line and X-ray properties}
\centering
\label{lines}
\begin{minipage}[240mm]{\columnwidth}
\begin{tabular}{lccccccccccccc}
\hline
Object &
z &
$\log{ L_{2-10}} $ &
$N_H$ &
$\Gamma_{2-10}$ &
\Lya\footnote{All line intensities are in units of $10^{-18}\,\,{\rm erg}\,{\rm s}^{-1}\,{\rm cm}^{-2}$\\
$\Gamma=1.80$ or $\Gamma=1.90$ means a guess of the 2--10 keV slope.\\
Column density in units of $10^{22}\,\,{\rm cm}^{-2}$\\
X-ray luminosities are corrected for intrinsic absorption.\\
}&
\civ\ &
\heii\ &
\CIIIb\ &
\MgII\ &
\oii\ &
\neiii\ &
\hb\ &
\oiii\ \\
\hline
CDF-S-10 & 0.424 & 42.66 &    1.5 & 1.17 &           &      &           &      &    &  43.0 &       &           &               \\
CDF-S-18 & 0.979 & 44.05 &    1.9 & 1.74 &           &      &           &      &    &  82.1 &       &           &               \\
CDF-S-20 & 1.016 & 43.27 &    5.6 & 1.78 &           &      &           &      &    &  21.4 &       &           &               \\
CDF-S-27 & 3.064 & 44.33 &   28.1 & 1.22 &  12       &  6.2 &           &      &    &       &       &           &               \\
CDF-S-41 & 0.667 & 43.16 &    5.6 & 1.45 &           &      &           &      &    &  122  &  46.4 &           &  425          \\
CDF-S-43 & 0.737 & 42.87 &    1.7 & 1.43 &           &      &           &      &    &  9.18 &       &           &               \\
CDF-S-45 & 2.291 & 44.04 &    8.2 & 1.46 &  12.5     &      &           &      &    &       &       &           &               \\
CDF-S-47 & 0.733 & 43.02 &    8.0 & 1.80 &           &      &           &      &    &  22.7 &       &           &               \\
CDF-S-51 & 1.097 & 44.01 &   22.4 & 1.72 &           &      &           &      &    &  36.7 &       &           &               \\
CDF-S-54 & 2.561 & 43.94 &   10.7 & 1.38 &  24.9     &  5.19&           &      &    &       &       &           &               \\
CDF-S-56 & 0.605 & 43.31 &    1.6 & 1.25 &           &      &           &      &    &  347  &       &  190      &  841          \\
CDF-S-57 & 2.562 & 44.20 &   19.3 & 1.69 &  109       & 16.6&  5.66     &  12.2&    &       &       &           &               \\
CDF-S-66 & 0.574 & 43.20 &    6.6 & 1.46 &           &      &           &      &    &  34.6 &       &           &               \\
CDF-S-75 & 0.737 & 43.43 &    3.7 & 1.21 &            &      &          &      &    &       &       &           &  35.5         \\
CDF-S-76 & 2.394 & 44.39 &   15.4 & 1.66 &  11.66     &      &          &      &    &       &       &           &               \\
CDF-S-85 & 2.593 & 43.74 &    8.7 & 1.80 &  22.3      &      &          &      &    &       &       &           &               \\
CDF-S-112 & 2.940 & 44.06 &   29.0 & 1.80 &  50.8     & 12.1 &  6.38    &      &    &       &       &           &               \\
CDF-S-117 & 2.573 & 43.77 &    3.1 & 1.80 &  27.6     &      &           &     &    &       &       &           &               \\
CDF-S-132 & 0.908 & 42.60 &    2.4 & 1.80 &           &      &           &     &    &  6.97 &       &           &               \\
CDF-S-151 & 0.604 & 43.07 &   23.2 & 1.80 &           &      &           &     &    &  14.2 &       &           &               \\
CDF-S-155 & 0.545 & 42.27 &    3.6 & 1.80 &           &      &           &     &    &  66.1 &       &  45.2     &  333          \\
CDF-S-176 & 0.786 & 42.95 &    2.2 & 1.80 &           &      &           &     &    &  16.4 &       &           &               \\
CDF-S-188 & 0.734 & 42.15 &    4.4 & 1.80 &           &      &           &     &    &  8.44 &       &           &               \\
CDF-S-189 & 0.755 & 42.60 &    7.5 & 1.80 &           &      &           &     &    &  5.44 &       &           &               \\
CDF-S-190 & 0.733 & 43.01 &   12.5 & 1.80 &           &      &           &     &    &  125  &       &           &  176          \\
CDF-S-201 & 0.679 & 42.59 &    2.6 & 1.80 &           &      &           &     &    &  25.7 &  8.63 &           &  114          \\
CDF-S-204 & 1.223 & 42.43 &    7.5 & 1.80 &           &      &           &     &    &  18.4 &       &           &               \\
CDF-S-252 & 1.172 & 43.21 &   15.8 & 1.80 &           &      &           &     &    &  29.6 &       &           &               \\
CDF-S-260 & 1.043 & 43.01 &   36.7 & 1.80 &           &      &           &     &    &  21.4 &       &           &               \\
CDF-S-264 & 1.316 & 43.23 &   21.6 & 1.80 &           &      &           &     &    &  16.8 &       &           &               \\
CDF-S-266 & 0.735 & 43.33 &   88.8 & 1.80 &           &      &           &     &    &  38.5 &       &           &  97.6        \\
CDF-S-267 & 0.720 & 43.17 &   14.2 & 1.80 &           &      &           &     &5.97&       &       &           &               \\
CDF-S-268 & 1.222 & 44.10 &   80.4 & 1.80 &           &      &           & 13.1&    &       &       &           &               \\
CDF-S-516 & 0.667 & 42.14 &    2.8 & 1.80 &           &      &           &     &    &  39.0 &       &           &               \\
CDF-S-519 & 1.034 & 42.41 &    1.1 & 1.80 &           &      &           &     &    &  17.7 &       &           &                \\
CDF-S-534 & 0.676 & 42.27 &    6.6 & 1.80 &           &      &           &     &    &  16.0 &       &           &                \\
CDF-S-535 & 0.575 & 42.19 &    2.9 & 1.80 &           &      &           &     &    &  50.0 &       &           &                \\
CDF-S-547 & 2.316 & 44.03 &   56.9 & 1.80 &  17.9 &          &           &     &    &       &       &           &                \\
CDF-S-580 & 0.664 & 42.09 &   10.5 & 1.80 &           &      &           &     &    &       &       &           &  43.7          \\
CDF-S-585 & 1.212 & 42.56 &    1.5 & 1.80 &           &      &           &     &    &  13.8 &       &           &                 \\
CDF-S-606 & 1.037 & 42.94 &   18.8 & 1.80 &           &      &           &     &    &  9.39 &       &           &                 \\
CDF-S-611 & 0.979 & 43.20 &   62.3 & 1.80 &           &      &           &     &    &  11.2 &       &           &                 \\
CDF-S-612 & 0.736 & 43.08 &   63.3 & 1.80 &           &      &           &     &    &  23.7 &       &           &                \\
CDF-S-615 & 0.759 & 42.17 &    7.4 & 1.80 &           &      &           &     &    &  11.7 &       &           &                 \\
CDF-S-633 & 1.374 & 43.72 &   86.7 & 1.80 &           &      &           & 31.3&18.0&       &       &           &               \\
H05370043 & 1.797 & 44.77 &   10.5 & 1.90 &           &  119 &           &     &    &       &       &           &                  \\
H05370164 & 1.824 & 44.48 &0.0\footnote{{X-ray spectral parameters not available, classification based on optical spectrum}} & 1.90 &           &  16.6 &          &     &    &       &       &           &                 \\
H05370016 & 0.995 & 44.35 &    1.3 & 1.90 &           &       &          &     &    &  91.8 &       &           &                 \\
H05370123 & 1.153 & 44.72 &    6.6 & 1.90 &           &       &          & 51.6&    &  90.5 &       &           &                 \\
H15800062 & 1.568 & 44.80 &   26.3 & 1.90 &           &  51.6 &          & 26.8&    &       &       &           &                 \\
H15800019 & 1.957 & 44.84 &    7.3 & 1.90 &           &  451 &           &     &    &       &       &           &                 \\
H50900013 & 1.261 & 44.55 &    2.5 & 1.90 &           &       &          & 31.5&    &  70.1 &       &           &                 \\
\hline
\end{tabular}
\end{minipage}
\end{table}
\normalsize
\twocolumn

The uncertainties on the X-ray measured fluxes can be obtained from Tozzi et al. (2006).
They are of the same order or smaller than the uncertainties associated with
X-ray variability (about a factor 2). The uncertainties on the emission line measurements
are typically 25\%. The combined uncertainty on the line to X-ray continuum, used later
in our work, are therefore of order 2.

 We have increased our
sample size by including seven type-II QSOs from the spectroscopic
follow-up of the HELLAS2XMM survey\footnote{The optical spectra are publicly available at:
http://www.bo.astro.it/$\sim$hellas/sample.html}. 
The X-ray spectral analysis for the additional sources
 has been performed by Perola et al.
(2004) and the X-ray selection criteria fulfill those  applied to
the CDF-S sample.  The emission lines of these sources have been measured in a similar way to
the other sample.
Thus, the new X-ray selected sample includes 52 type-II AGNs out of which 18 are classified
as type-II QSOs.
 All fluxes were converted to
luminosities assuming a $\Lambda$-cosmology with $\Omega_m=0.3$,
$\Omega_{\Lambda} = 0.7$ and H$_0 = 70$ km s$^{-1}$ Mpc$^{-1}$.

\subsection{Comparison with earlier type-II samples}

In order to compare our new results  with previous works of this type, we made extensive use of the data in
M94, A97, Polletta et al. (1996) and Bassani et al. (1999). These references include X-ray and \oiii\ 
fluxes for about 60 type-II mostly low X-ray luminosity, low redshift sources.
Unfortunately, some of the X-ray data in M94 and in A97  
were obtained with pre-ASCA instruments and the procedures used to obtain the 2--10 keV absorption
corrected luminosities suffer from various uncertainties. Moreover, the quality of the data did not
allow the clear identification of Compton thick sources that are likely to bias the \oiii-X-ray comparison.
Given this, we prefer to use the more uniform, carefully measured
and modeled X-ray observations in Turner et al. (1997; 1998) and in Bassani et al. (1999). 
These are  based on ASCA and BeppoSAX observations and
considered to be more reliable. For example, a comparison of the Turner et al. results (see Table 12 in 
Turner et al. 1997) with M94 shows a systematic trend for larger L(2--10 keV) luminosities in the
latter, especially for low luminosity X-ray sources. 
As shown below, this has important consequences  to the  main conclusion of M94.

The improved Turner et al. (1997) and Bassani et al. (1999) absorption corrected 2--10 keV 
luminosities were combined with the \oiii\ fluxes listed
in M94 and in Polletta et al. (1996) to obtain a high quality sample of low
redshift type-II sources. Since there is much overlap between those lists, we chose as our primary source
the Bassani et al. (1999) sample but carried a similar statistical analysis also for the
Turner et al. data set with basically identical results.
As explained earlier, we omit several Compton thick sources found in this sample. 
 We have also excluded all objects with a noticeable Seyfert 1 contribution
(we only include source listed as S1.8, S1.9 and S2), two AXJ sources (see
Bassani et al. for justification) and sources with $z>0.05$. The result
is a sample of 42 low redshift type-II AGNs. The cut in redshift is only to enable a clearer 
comparison with the new, high redshift sample.
We have verified that the inclusion of the 6 sources omitted by the redshift
criterion changes nothing for the statistical analysis that follows.
For comparison we note that the original M94 sample of
type-II sources introducing the idea of a constant  \LOLX, 
includes only 16 low redshift objects.

The new high redshift sample requires a different procedure since it includes
sources with various emission lines, mostly \oii, and only 8 measurement of the \oiii\ line intensity.
However, there is enough information in the literature to enables the conversion
from all those lines to \oiii\ intensity. We used a variety
of data sets where UV as well as \oiii\ line fluxes are available. For \La\ in low redshift
Seyfert 2s, we used ground-based observations of \oiii\ and IUE,  HST and FUSE observations
of the UV lines. Such data are very limited and the most useful sources  
are Ferland and Osterbrock (1986) and 
Kuraszkiewicz et al. (2004).
For this sample we find a mean of 0.81 and a median of 0.60 for L(\Lya)/L(\oiii).
There are also several extensive studies of narrow emission line galaxies,
mostly radio loud high redshift galaxies,
where rest-frame UV lines were observed from the ground, in the visual band, and
the \oiii\ line was observed in the near IR, in the H, J or K-bands. Work of this type is 
discussed in Evans (1998), Vernet et al. (2001), Stern et al. (1999), Rottgering et al. (1997),
Stern et al. (2002) and Norman et al. (2002). 
Out of these references we found 11 objects that were considered appropriate to investigate 
the I(\Lya)/I(\oiii) ratio in narrow line active galaxies. We found 
a mean of 2.40 and a median of 1.36 for the above line ratio in this group.
Some of the new sources do not have measured \Lya\ and we had to use a similar method based on
different UV lines, e.g. \civ\ or \CIIIb. 
All samples used to define these  ratios are small and cover a large range in the properties 
under study. Therefore we prefer  to use medians rather than 
means and a ``typical range'' which includes 2/3 of the objects.

Our new sample contains 34 sources with measured  \oii\ line flux, thus the conversion to L(\oiii) in
this case is of a different nature since the final analysis will be dominated by those sources.
We used the largest available sample where \oii\
and \oiii\ luminosities are directly compared (291 sources in Zakamska et al. 2003). The mean 
L(\oiii)/L(\oii) in this sample is somewhat luminosity dependent and is close to
3.5 over the range of L(\oii) found here. This is also close to the median of the 
6 sources in our sample showing both \oii\ and \oiii\ lines.
Table 2 summarizes all these measurements
by listing a composite type-II spectrum that was used to
define an ``effective'' \oiii\ intensity for our high redshift sources. Many other composite spectra
have appeared in the literature, e.g. Ferland and Osterbrock (1986) and Zakamska
et al. (2003).
The most important results that are relevant here are that the median value of I(\Lya)/I(\oiii)
is very close to 1.0 and the one for L(\oiii)/L(\oii) is about 3.5.
\begin{table}
\caption{Composite type-II AGN spectrum}
\label{composite}
\centering
\begin{tabular}{lcc}
\hline\hline
Line &
relative intensity  &
range \\
\hline
\oiii\  & 1.0 & 1.0 \\
\Lya\   & 1.0 & 0.6--1.6 \\
\civ\   & 0.30& 0.2--0.5 \\
\heii   & 0.15& 0.1--0.4 \\
\CIIIb\ & 0.2 & 0.1--0.4 \\
\oii\   & 0.29 & 0.2--0.8 \\
\hline
\end{tabular}
\end{table}

Having defined an effective L(\oiii) from the observations of other emission lines,
we can now compute \LOLX\
for all the new X-ray sources and compare them with the low redshift samples. This
is shown in Fig. 1 where we display such ratios for the 52 new sources and
also mark those where \LOLX\ is based on
a conversion of a measured L(\oii).
 As seen from the diagram, there is a systematic trend in a sense that \LOLX\
is considerably smaller for the higher X-ray luminosity sources. 
The statistical analysis confirms the high level of significance of the correlation. The Spearman
rank correlation coefficient for the 52 sources is $r_s=0.66$ ($p=1.0 \times 10^{-7}$) and a
simple linear regression gives 
%
\begin{figure}
\centering
  \includegraphics[width=9cm]{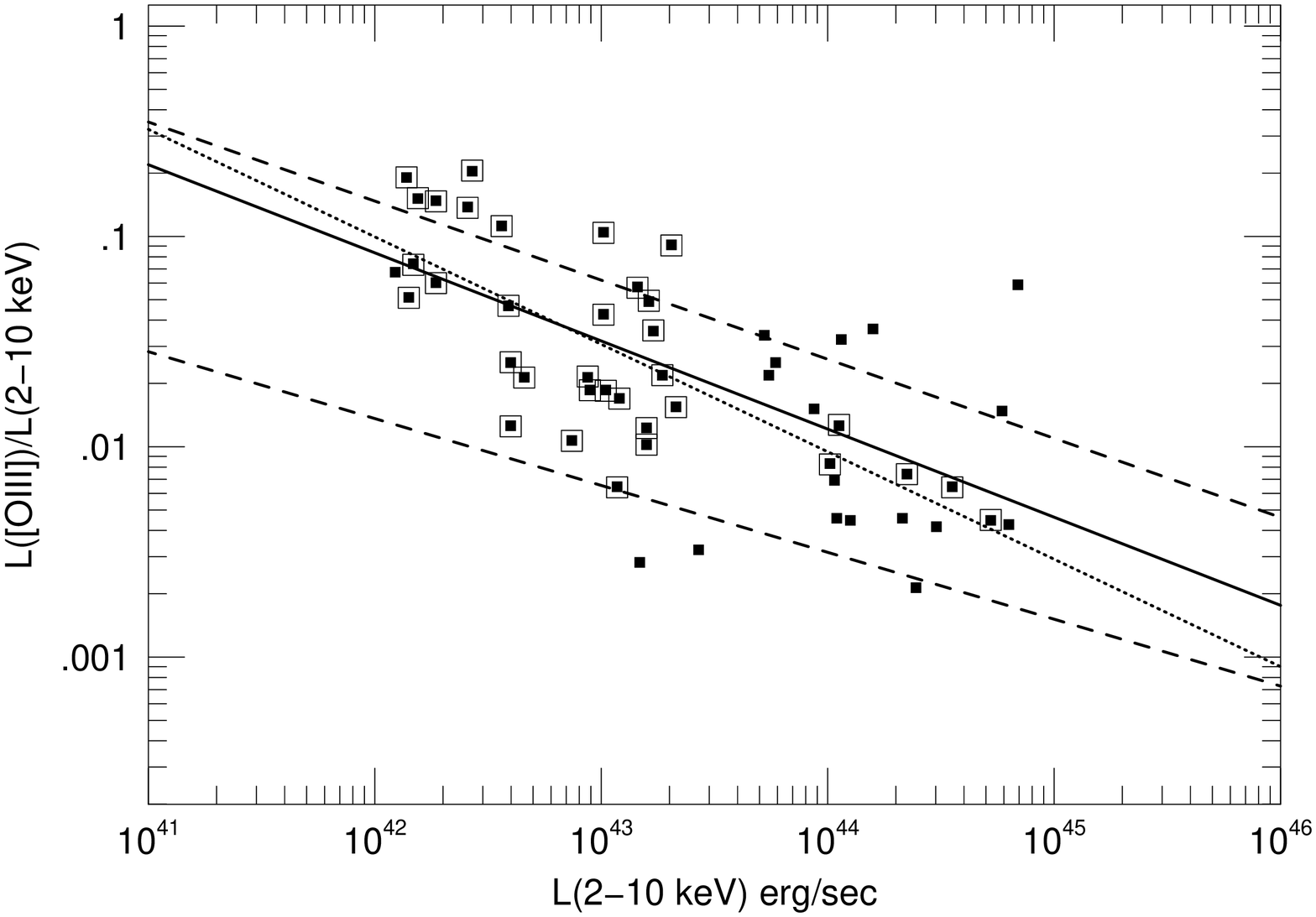}
\caption
{\LOLX\  vs. L(2--10 keV) for X-ray selected type-II AGNs. Full squares are
data for the new high redshift sources using the ``effective'' \oiii\ luminosity
as described in the text.
Open squares are sources with measureable \oii\ line.
The solid line is a fit to the entire sample and
the dotted line a fit to the \oii\ sample only.
Dashed lines are best fits to the Bassani et al. (1999) modified sample (see text).
Lower curve: observed fluxes. Upper curve: line fluxes corrected for reddening.
}
\label{typeII-oiii}
\end{figure}
%
%
\begin{equation}
 \begin{array}{l}
\log{  \frac{ {\rm L}_{[OIII]}}{ {\rm L}_{2-10} } }{\rm (type\,II\,\,uncorrected)}  \\
 \,\,\,\,\,\,\,\, = (16.5 \pm 2.9) -(0.42 \pm 0.07 ) \log{ {\rm L}_{2-10} } \,\, ,
 \end{array}
\end{equation}
where we use ``uncorrected'' to indicate that the emission lines are not corrected for reddening.
Since so many measurements are based on one line, we have also tested the correlation in the sub-sample of
34 sources with \oii\ emission line. This also gives a very strong correlation
 ($r_s=-0.70\,\, p=4.9\times 10^{-6}$)
with a somewhat steeper slope of $-(0.51 \pm 0.09)$. The slopes of the 
correlations are consistent within the errors.
As evident from the diagram, the scatter is larger than the estimated uncertainties on
individual points suggesting that there is a large range in the intrinsic properties of
the sources.
Finally we tested the correlation for the sub-sample of 24 sources showing emission lines
other than  \oii.
This sub-sample shows a weaker but significant correlation ($r_s=-0.57 \,\, , p=0.003$) 
with a slope of -$(0.39 \pm 0.14)$. For completion we note that in our new sample,
L(2--10 keV)$\propto$L(\oiii)$^{0.58}$, i.e. similar to the typical correlation between X-ray
and UV continuum luminosities (e.g. Strateva et al. 2005). 

Next we tested the 42 low redshift sources obtained from the Bassani et al. (1999) sample. The
statistical analysis shows a weaker but significant correlation 
($r_s=-0.44$, $p=3.96 \times 10^{-3}$)
 and a slope that is somewhat flatter ($-0.32 \pm 0.1$) than the one found in the new sample.
The best regression line is shown in Fig. 1.
However, the mean \LOLX\ in this sample is considerably lower than the corresponding mean
in the new sample (a factor of about 4  at L(2--10 keV)=$10^{44}$\ergsec). 
As discussed later, we suspect that the difference is due to the way the sample was selected and the large
amount of reddening (see \S3.2).
We also checked the old low redshift data obtained from M94, A97 and Turner et al. (1997) using
only the more reliable X-ray data in Turner et al. The \LOLX\ vs. L(2--10 keV) correlation is already present
in this data set with a form very similar to what was found here.
Thus, those correlations were present, yet never noticed, in older type-II AGN samples.
 
We note in passing two other sources of data for high luminosity type-II sources.
Zakamska et al. (2004) identified six ROSAT all sky survey (RASS)
 type-II sources with measured \oiii\ and X-ray fluxes.
As explained in their paper, the estimated 2--10 keV flux is highly uncertain because of the limited
ROSAT response at those energies and the likelihood of much absorption at low X-ray energies.
Zakamska et al. suggest that the values of $L_X$ given in Table 4 of their paper
are conservative lower bounds  to the intrinsic 2--10 keV luminosities.
Using those numbers we could obtain a conservative upper bounds of 0.013 for the median of
\LOLX\  in this sample. This places the sources in the same region of Fig. 1 as our new objects with
comparable X-ray luminosities. We
prefer not to include these uncerain numbers in our statistical analysis.
Ptak et al. (2005) list six new X-ray measurements for type-II SDSS sources. The \LOLX\
in this small groups shows a large scatter with a mean which is larger than the one found
in our X-ray selected sample. As discussed below, this is likely to be attributed to
a systematic difference between X-ray selected and optically selected samples and hence we
do not include those AGNs in our analysis.

\subsection{\LOLX\ in type-I AGN}
To complete the description, we have constructed a similar ratio for type-I
AGNs. There are many references for L(\oiii) but fewer for  L(2--10 keV). Some are already
included in M94 and A97 who, at the time, did not have data on  very high luminosity
sources.
There are also several detailed studies of the X-ray properties
of low redshift quasars with measured L(\oiii). These are mostly PG sources that were
observed by ASCA and BeppoSAX and more recently by \chandra\ and \xmm. Some of the data are given in 
George et al. (1998), Laor et al. (1998), Kaspi et al. (2005),
Piconcelli et al. (2005) and Heckman et al. (2005). The most
comprehensive of those is the Piconcelli et al. (2005) \xmm\ data set.
We note
that corrections for intrinsic absorption at the 2--10 keV range 
are very small in most of those sources except for a few (e.g. NGC\,3783) that
are known to have large column warm absorbers. Thus the main source of scatter is the  intrinsic 
X-ray variability.
We also note that the
X-ray fluxes in Heckman et al. (2005) require a correction factor of 1.35 to take into account the 
somewhat different energy band (3--20 keV).
The final sample contains 68 sources that are plotted in Fig. 2.
\begin{figure}
\centering
  \includegraphics[width=9cm]{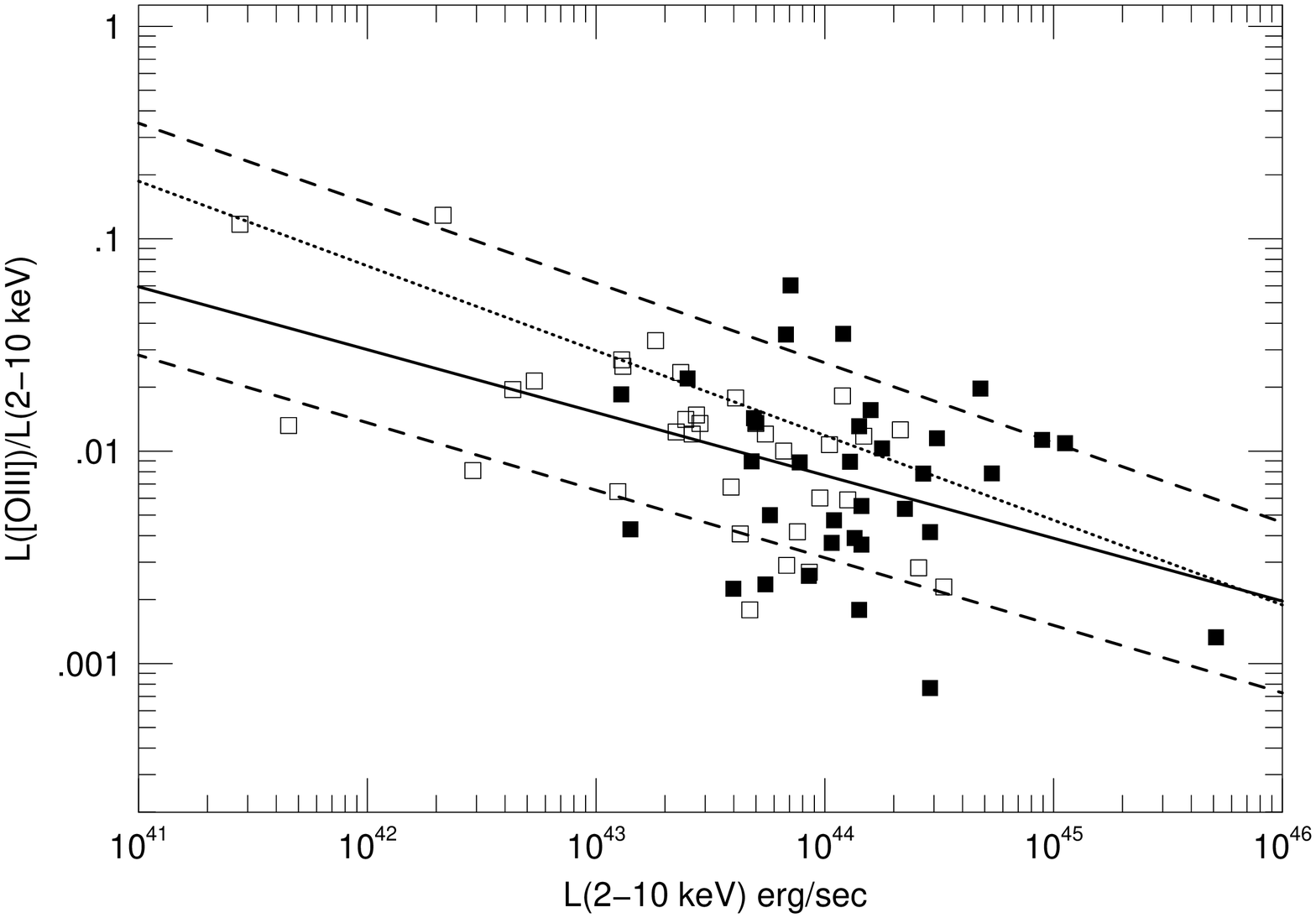}
\caption
{Same as figure 1 but for type-I sources.
Open symbols:
data from M94,  A97 and Heckman et al. (2005).
Full squares: data from BG92 and Piconcelli et al. (2005).
The solid line is a linear fit to the data and the dotted line the fit to the type-II sample.
Lower and upper dashed lines are the Bassani et al. (1999) curves shown in Fig. 1.
}
\label{typeI-oiii}
\end{figure}

The statistical analysis of the type-I sample shows a strong correlation with 
 $r_s=-0.43$ ($ p=2.4 \times 10^{-4}$). 
Again, this correlation was missed in the earlier
M94 and A97 works. The linear regression result is
\begin{equation}
 \begin{array}{l}
\log{  \frac{ {\rm L}_{[OIII]}}{ {\rm L}_{2-10} } }{\rm (type\,I\,\,uncorrected)} \\
  \,\,\,\, = (10.9 \pm 2.8) -(0.296 \pm 0.06 )  \log{ {\rm L}_{2-10} } \,\,.
 \end{array}
\end{equation}
This relationship is plotted as a solid line in the diagram. We also show, as a dotted line, the
relationship found for type-II sources as well as the two curves for the Bassani et al. (1999)
type-II sample. As seen, the slopes of the correlations for the 
X-ray selected type-II sample and the optically selected type-I sample are
are somewhat different, but are consistent within $2 \sigma$.
As argued in \S3, this does not mean a similar behavior of type-I and type-II
sources with regard to their optical vs. X-ray properties.
 For completeness, we have analyzed the Heckman et al. (2005) sample in the same way. We 
used the 34 type-I X-ray selected sources listed in this paper and found a very similar correlation to
the one presented here. Heckman et al. discuss the linear relationship between L(2--10 keV) and L(\oiii) but
 do not comment on the luminosity dependence of \LOLX.
We must also note that L(\oiii)/L(\oii) in type-I sources is larger than in type-II sources (e.g.
Zakamska et al. 2003; see also the special case of radio-loud sources in Jackson and Brown 1990).
This may introduce a systematic uncertainty when comparing our new sample (basically \oii\ lines) with
optically selected sample with measured \oiii.

\section{Discussion}

Our new X-ray selected sample of 52 high redshift  
sources is the first of this type which is large enough to allow meaningful statistical analysis
of the line-to-X-ray properties of
type-II sources over a large luminosity range.
The major findings of this work are strong dependences of \LOLX\ and \LOIILX\ on L(2--10 keV), in type-II
AGNs, contrary to past claims.
Moreover, re-analysis of older data sets that do not include high luminosity, high redshift
 sources, also show a similar correlation. 
Our work shows a correlation of those properties in type-I AGNs, also
in contrast to earlier claims.

The present analysis, as well as similar past works, is subjected to various uncertainties
and biases. The two most important ones are related to sample selection and
to narrow emission line reddening in AGNs.

\subsection{X-ray selected vs. optically selected AGN samples}

High quality X-ray spectroscopy of type-II sources with small to moderate obscuring columns can
be used to reliably recover the intrinsic L(2--10 keV) of the source. The only 
uncertainty is intrinsic continuum variablity which
introduces a scatter of a factor $\sim 2$. This should not affect the correlation in 
large samples.
More important is the difference between X-ray selected and optically selected samples.
As discussed in many papers, most recently in Heckman et al. (2005), X-ray samples are
biased towards X-ray bright sources and hence, in the case under study, will  produce a
smaller mean \LOLX\ compared to optically selected ones. While the Heckman et al. (2005) sample seems
to confirm this suggestion, the data shown in Fig. 1 here seems to be in conflict with it since the
mean \LOLX\ in our X-ray selected sample is larger than in the Bassani et al. (1999) sample.
We suggest two reasons for this difference.
One is line reddening which is discussed in \S3.2 and the other is the difference
between optical selection which is based on emission line strength (i.e. EW) and selection based on 
optical continuum flux.

Older samples of type-II sources were selected in various different ways. Some were found via direct
spectroscopy of nearby bright emission line galaxies and others due to various other properties. This is
clearly the case for most sources in M94, A97, Bassani et al. (1999)
and other older AGN samples. The SDSS is an $i$-mag selected sample and is hence biased towards high
continuum flux type-I AGNs and strong emission line type-II AGNs. The former will be biased against large EW
lines
while the later are expected to be
biased against weak emission line sources. Thus, there are three possible biases to consider, two in 
optically selected samples and one in X-ray selected samples.
Past, non-uniformally selected type-II sources that were found mostly in bright galaxies 
 are not necessarily expected to show stronger
\LOLX\ compared with sources in 
our new  X-ray selected sample. On the other hand, we expect X-ray follow up on type-II SDSS sources
to show larger \LOLX. Given this, the comparison of
past, randomly selected type-I and type-II samples is, perhaps, meaningful since many of
those sources were discovered by the same techniques  and are typical of optical continuum selected sources. 
More examples related to type-I SDSS sources are presented in \S3.3.

\subsection{Emission line reddening}

Line reddening in type-II sources depends on the geometry and location of the NLR.
A good example of the location dependent extinction
is found in the recent work of Collins et a. (2005) on Mrk 3. Other detailed studies based
on HST spectroscopy can be found in the literature (e.g. Ferruit et al 1999). These few HST-based
studies, combined with the IUE-based analysis of Ferland and Osterbrock (1986), suggest that the narrow line
reddening corresponds to  E(B-V) in the range 0.2--0.4 mag. Other methods, based on the
observed Balmer decrement, can be used to obtain the volume averaged narrow line reddening.
For example, the median reddening found by Dahari and De-Robertis (1988) in
type-II sources is E(B-V)$\sim 0.54$ mag.
These authors  also mention that the (less certain) line reddening in type-I sources is smaller, by about 0.1--0.2 mag.
Bassani et al (1999), following Maiolino et al. (1998), used various references to obtain \Ha/\Hb\
for all their sources.  The
median reddening in their sample is significantly larger, corresponding to E(B-V)$\sim 0.65$ mag.
More and independent confirmation of the large \Ha/\Hb\ in some of
those Seyfert 2 galaxies can be found in Storchi-Bergmann, Kinney and Challis (1995).
All those measurements give 
larger E(B-V) than the  mean E(B-V)$\sim 0.27$ mag. quoted
by Zakamska et al. (2003) for their type-II SDSS AGN sample.
The difference may be another manifestation of the bias against faint line 
type-II AGNs in the SDSS sample.
This is likely to be more important at z$>0.2$, where the type-II SDSS sample is incomplete.

To quantify the effect of reddening, 
we have used the Bassani et al. (1999) lists to obtain the reddening 
corrected \LOLX\ vs. L(2--10 keV) for the 42
sources discussed in \S2. We assumed an intrinsic \Ha/\Hb=3.1 (e.g. Ferland and Osterbrock 1986) and obtained 
\begin{equation}
 \begin{array}{l}
\log{ \frac{ {\rm L}_{[OIII]}}{ {\rm L}_{2-10} } }{\rm(type\,II\,\,corrected)} \\
 \,\,\,\,  = (15.0 \pm 4.0) -(0.38 \pm 0.09 )  \log{ {\rm L}_{2-10}  }\,\, ,
 \end{array}
\end{equation}
i.e. a significant correlation with a somewhat steeper slope than the one with no reddening. The shift in the mean
\LOLX\ between the reddening corrected and the observed correlations, 
at L(2--10 keV)=$10^{44}$ \ergsec,  
is a factor of  $\sim 8$. 
There is no tendency for more luminous X-ray sources to show more line reddening (which could have been
an explanation for the decreasing \LOLX\ with the X-ray luminosity).

Our new data contain mostly  UV lines and we have no way to estimate reddening correction factors
for individual sources. Regarding the mean properties of the sample, we
note that the expected intensity ratio of I(\Lya)/I(\oiii) is $\sim 9$. This number is obtained from combining
the theoretical I(\Lya)/(\Hb)$\simeq 50$ (e.g. Netzer 1982; Ferland and Osterbrock 1986) with the observed
I(\oiii)/I(\Hb)$\simeq 6$ (e.g. Zakamska et al 2003 table 2).
 This is a factor $\sim 9$ larger than the empirical ratio  (Table 2) presented
here and can be translated (assuming line of sight reddening)
to galactic E(B-V)$\sim 0.3$ mag. Obviously, this average value is not very meaningful
 given the large scatter  in I(\Lya)/I(\oiii).

A related issue is the different mean \LOLX\ found between 
optically selected type-I and type-II samples.
The optically selected samples studied here, albeit small and incomplete, show a clear shift in a
sense that L(\oiii) in type-I sources is larger, by a factor of $\sim 2$, for the same L(2--10 keV). 
A possible explanation for the shift is additional reddening of \oiii\ in type-II sources. This
can be the case if some of the reddening is due to a large structure, like the inner galactic
disk, in disk-dominated systems with different orientations to the line-of-sight.

The results found here may look in conflict with studies of I(\oiii)/I(\oii) in type-I
and type-II sources (Zakamska et al. 2003 and references therein). Such studies  show this ratio
to be larger in type-I sources despite the shorter wavelength of \oii, implying perhaps more reddening. 
However, the \oii\ line emission
region is much larger than the \oiii\ region and we do not expect
the same amount of reddening in those two parts of the NLR. 
Other differences between type-I and type-II narrow line properties have been noted in the past (e.g.
Zakamska et al. 2003 and references therein). Some of those may be related to similar geometrical
factors. The  conclusion is
 that the \oii\ line luminosity may be a better measure of 
the NLR intrinsic emission and use of the \oiii\ flux should be treated with care.

 Finally we comment on the fact that the \oiii\ emission is known to depend on radio properties being
stronger in radio-loud sources. We have 20 cm radio VLA data for all sources in our sample
(Kellerman et al. 2006, in preparation). Only
3 were detected to a flux limit of 42$\mu$Jy. Thus, radio properties cannot
influence much the present results.

\subsection{L(\oiii) as a luminosity indicator in AGNs}
The correlations found here allow us to investigate the  use of L(\oiii) as
a luminosity indicator in AGNs. For this we look for the correlation of the
equivalent width of this line with the 
optical continuum, $L_{5100}$,  defined here as the 
uncorrected $\lambda L_{\lambda}$ at 5100\AA. 
To obtain an expression for the expected EW(\oiii) we assume a  0.1--1 $\mu$m power-law continuum 
with an optical energy slope of $\alpha_{op}=0.5$ and a hard X-ray (2--10 keV) energy slope of 
$\alpha_x=0.9$. Given this spectral energy distribution (SED), and 
$\alpha_{ox}$ (the energy slope connecting
2500\AA\ and 2 keV), we find 
$\log{ (L_{2-10}/L_{5100}) }=3-2.605 \alpha_{ox}$. 
We can thus write the following expression for EW(\oiii),
\begin{equation}
\log {( {\rm EW([OIII]])}) } = 6.7-2.605  \alpha_{ox} + 
 \log{  \frac{ {\rm L}_{[OIII]}}{ {\rm L}_{2-10} } } \,\, , 
\end{equation}
which again assumes no reddening. A specific EW(\oiii) can be obtained by taking into account
the known dependence of $\alpha_{ox}$ on UV and optical luminosity (Strateva et al. 2005).

Data on EW(\oiii) in very large samples are now available from the 2dF and the SDSS samples. This line is
relatively easy to measure in luminous type-I sources and work of this type has been published in several
papers (e.g. S05; Hao et al 2005). We have made our own study of the SDSS archive using basically
all type-I sources available at this stage with z$<0.75$. 
A detailed description of the procedure used to measure all emission lines 
will be given elsewhere (Netzer and Trakhtenbrot, 2006). In brief, we have used the fourth data release
(Adelman-McCarthy et al. 2006)
and included only sources within the magnitude
limit of this sample (i=19.1 mag.).
We measured the \oiii\ line flux and EW by using {\it single
Gaussian} fits to the two ${\rm [OIII]}$ lines and by employing a sophisticated procedure to deblend and subtract
the strong FeII lines over the 4400--5500\AA\ range.
This underestimates, somewhat, the real \oiii\ intensity since the  line is known to have
an extended blue wing in many AGNs 
(Greene \& Ho  2005 and references therein). However, the single-component fit is more robust, especially
in type-I sources because of the easier de-blending with the broad \hb\ profile. It also provides a  more
consistent way for comparison with type-II SDSS sources  that were mostly measured
in a similar way.
Some 2000 sources were removed from the analysis mostly because they did not fulfill the
AGN detection criteria and because of observational uncertainties
(see Netzer and Trakhtenbrot 2006 for more details). We chose not to remove those $\sim10$\% of
the sources that are radio loud but verified that none of the conclusions discussed below are
affected by their inclusion.
The end result is a sample of about 12,000 sources with measured intensities or upper limits
on the \oiii\ line.

The SDSS is a flux limited sample and hence suffers from various selection effects, especially
near its flux limit.
As explained earlier, this may affect the EW distribution of \oiii\ and other emission lines since
source selection is based on continuum SED (or ``colors'') and continuum flux (or broad
band magnitudes).
Because of this, we have investigated, separately, the luminosity dependent and the redshift
dependent distributions of EW(\oiii). 
The results are shown in Fig. 3 where the median EW(\oiii) is plotted in various ways.
 The diagram shows a tendency for 
EW(\oiii) to decrease as a function of $L_{5100}$  at z$>0.3$ but 
there is no obvious global redshift dependence. Adding all  redshift
bins gives a change of the median EW by a factor of about 1.7 from 
$L_{5100}=10^{43}$ \ergsec\ to $L_{5100}=10^{46}$ \ergsec.
The mean EW(\oiii) (not shown here) has a similar trend with $L_{5100}$ and is larger than the median by a factor of about 1.5.
\begin{figure}
\centering
  \includegraphics[width=9cm]{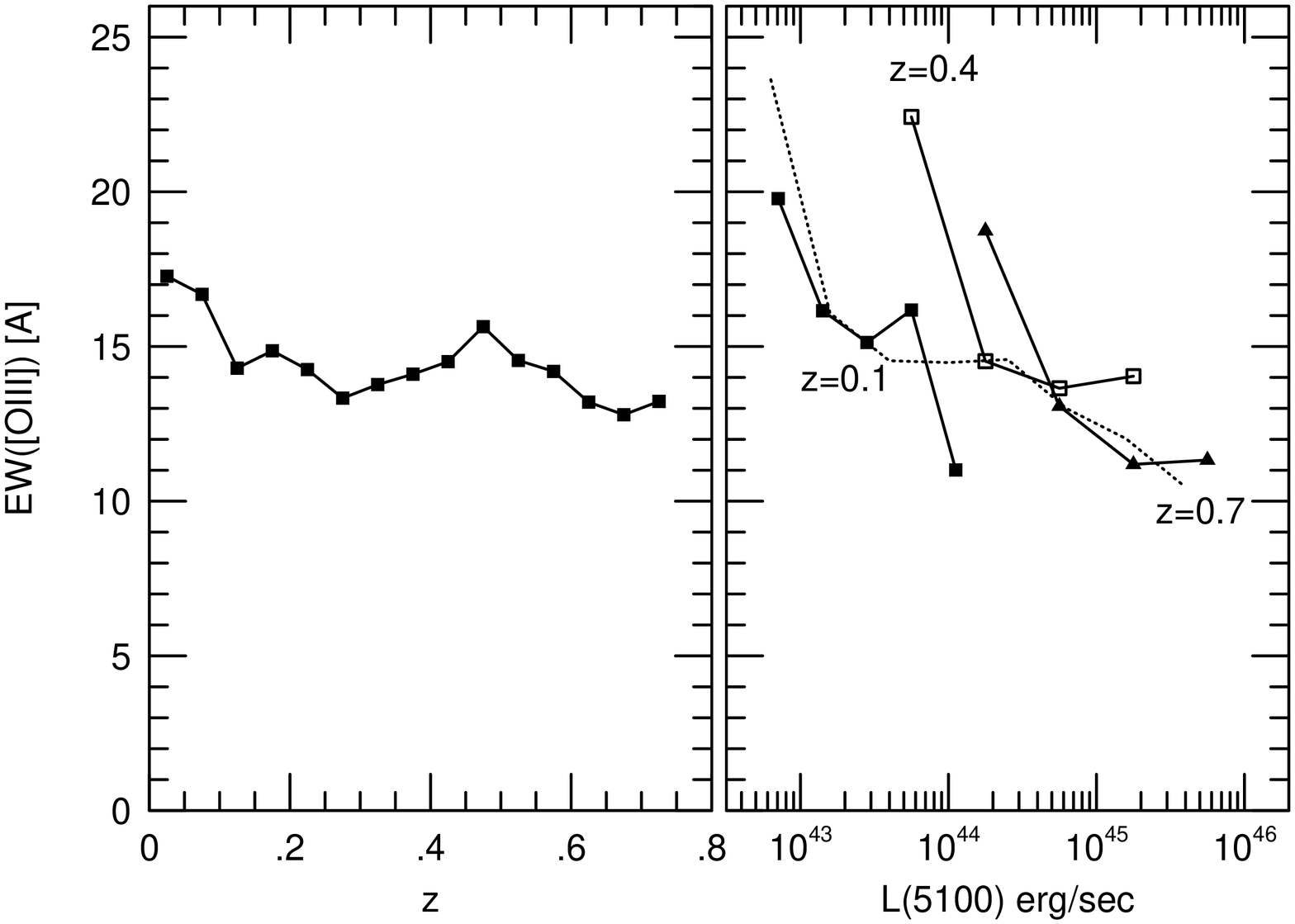}
\caption
{The median EW(\oiii) as a function of redshift (left) and continuum luminosity ($\lambda L_{\lambda}$ at 5100\AA, right)
for  11,900 type-I SDSS AGNs. The luminosity dependence is shown for three
different redshift bands (solid lines with symbols) and the entire sample (dotted line).
}
\label{EW_median}
\end{figure}

The tendency for EW(\oiii) to decrease with continuum luminosity (the
``Baldwin effect'', see  Baldwin 1977) has  been studied in previous work including in
Croom et al. (2002) and in Netzer et al. (2004). 
The Croom et al. study includes several thousands 2dF AGNs with measured \oiii\ line in
the redshift range 0--0.5.
The sample includes sources that are almost a magnitude fainter than the faintest SDSS AGNs.
According to this paper, there is no Baldwin relationship for this line. In fact, the line EW tends to
increase somewhat with increasing luminosity.
The Netzer et al. (2004) sample includes some of the most luminous high redshifts QSOs.
According to these authors, the Baldwin effect seen in previous  \oiii\ samples is not due to 
a decreasing mean line luminosity but rather
the larger fraction of high luminosity AGNs with extremely week, sometimes unobservable  \oiii\ lines. 
According to these authors, some very high luminosity AGNs show large EW(\oiii), similar to the one observed in lower luminosity sources,
but others show a 
very weak line. In such cases, the mean of the population may not be a very  
meaningful quantity since we may be dealing
with a bimodal distribution of equivalent widths.
In the Netzer et al. sample, the luminosity is more than an order of magnitude larger
than the highest luminosity considered in the present work and
about a third of the sources have undetectable \oiii\ lines. 

The analysis of the 12,000 type-I SDSS AGNs seem to be consistent with both Croom et al. (2002) and Netzer et al. (2004).
While there seem to be a trend in both median and mean at all redshifts (Fig. 3), this is not statistically
significant at the low luminosity, low redshift end.
This has been verified by a regression analysis of the lower redshift sources. We find that in the
range $0<z<z_1$, where $z_1<0.3$, there is no significant Baldwin effect while at higher redshifts, where the mean  L(\oiii) is
larger, there is a significant effect. The Croom et al. (2002) 2dF sample includes lower luminosity lower redshift sources
and the lack of a Baldwin effect in their data is
in agreement with our finding. The trend at larger luminosities and redshifts in our sample is statistically significant
and the mean EW(\oiii) is indeed decreasing with $L_{5100}$. Some of it, especially at the largest z considered here,
may be related to the flux limit of the SDSS sample.
Unfortunately,
we are not in a position to investigate in detail the relationship between EW(\oiii) and \LOLX\ in 
optically selected type-I sources
because of the small number of available L(2--10 keV) measurements. However, it is obvious that
\LOLX\ found here and shown in Fig. 1, drops faster with L(2--10 keV) compared with the drop of EW(\oiii) with $L_{5100}$
(note that the 
drop of $\alpha_{ox}$ with optical and UV continuum luminosity causes an opposite effect).

Given eqn. 4 we can work out the expected (i.e. as would be seen 
seen against an unobscured continuum) EW(\oiii) in our type I and 
type-II samples.  Assume 
$\alpha_{ox} =1.4$ and L(2--10 keV)=$10^{44}$ \ergsec\ (see Fig. 2). This would
give \LOLX$= 7.5 \times 10^{-3}$  (eqn. 2) and hence EW(\oiii)$\simeq 9$\AA,
close to the median at the high luminosity end in Fig. 3.
Thus, except for
an obvious difference between the median and the mean, the high luminosity type-I X-ray sources studied
here have typical values of EW(\oiii) similar to those  observed at the high luminosity end of our z$<0.75$ SDSS sample. 
For comparison, the Heckman et al. (2004) work on type-II sources assume a typical L(\oiii) which is
equivalent to EW(\oiii)$\simeq15$\AA.

Finally we comment on the 
expected number of high luminosity AGNs with detectable emission lines.
The mean (uncorrected)  L(\oiii)
observed at the highest luminosity end in Fig. 1 can be translated (see eqn. 4) to 
EW(\oiii)$<6$\AA.   At higher luminosities, like those found in the Netzer et al. (2004) sample, the
expected EW(\oiii) is even smaller and
EW(\oii) is about a factor of 5 below this number.
Such equivalent widths are close to the detection limit of many spectroscopic surveys.
These numbers suggest that many objects with
very high X-ray luminosities, would have extremely weak emission lines and would not be detected by  spectroscopic
follow ups of optically selected and X-ray selected samples (see the 
Netzer et al. (2004) more detailed discussion of this point).

A recent paper by Martinez-Sansigre (2005) discusses a sample of 21 type-II high luminosity AGNs 
discovered by their radio and infrared
properties in the Spitzer first look survey (SLF). Deep spectroscopic
follow-up resulted in narrow emission line detection of 10 of the sources. Eleven others, while being of similar
mid-IR properties, did not show a trace of any emission line. According to the paper, the sources without detected
narrow lines are those where galactic scale obscuration increased the line
reddening of the otherwise normal NLRs. We suggest that the lack of
detected narrow lines in AGNs found in such infrared surveys is
related to the present finding of extremely weak emission lines in X-ray detected AGNs. For example, 
the analysis of our own spectroscopic
follow-up (Szokoly et al. 2004)
shows that out of 106 type-II candidates selected by the column density and the X-ray luminosity
(based on photo-z) we could only obtain redshifts for 69. The remaining 37 sources (35\%) show no emission
lines. These
sources have X-ray luminosities similar to the ones presented  here and the integration times, and overall
observing conditions, were very similar to those used to find the sources listed in Table 1.

All the above findings are consistent with the Netzer et al. (2004) ``disappearing NLR''
suggestion that a large fraction of high luminosity AGNs of both types show
extremely weak narrow emission lines. One explanation is that in many such sources, the NLR size
exceeds the galactic size and most of the gas escapes the system.
Since the \oii\ emission region is considerably larger than the \oiii\ zone, we expect the effect to be
more pronounced in the former case, i.e. a steeper dependent of \LOIILX\ on L(2--10 keV) compared
with \LOLX, at the high luminosity end.
 Such extreme type-II sources will never be found by their emission lines. 
Another possibility, is extremely large
emission line reddening preferentially in the highest luminosity sources
but the data presented here do not support this view.
All this must be taken into account when assessing the type-II contribution to the CXB, using emission line
surveys.

\subsection{Implications to AGN luminosity function and space distribution}
AGN LFs have been the subject of much discussion in recent years, following the publication of several
large new systematic
surveys like  2dF and  SDSS. Such surveys  allow a comparison of various properties as a function
of source luminosity, redshift and even black hole mass. Comprehensive studies, and many references
can be found in Croom et al (2004), Hao et al. (2005) and S05.

Measured emission line fluxes in thousands of SDSS sources, 
allowed Hao et al. (2005a; 2005b) and S05 to construct emission line LFs for the \oiii\ line. The
line flux in these studies was assumed to be independent
of orientation and thus a good measure of the intrinsic source luminosity.
No reddening correction was used when deriving the LF.
Hao et al. (2005b) investigated sources with $0<z<0.15$ with the conclusion that the
two AGN types are indistiguisable at low luminosity but type-I sources outnumber type-II ones at
the high luminosity end. 
 S05 used a somewhat different redshift range (0.02--0.3) and claimed a more sever paucity  of type-II
sources at large \oiii\ luminosity.  S05 further compared the results with the prediction of the
``receding torus'' model. According to this model, the opening angle of the central
obscuring torus is luminosity dependent and higher luminosity sources have larger openings.
This suggests that the fraction of type-I sources would be larger at the high luminosity end.
More details of the model can be found in Lawrence (1991),
Simpson (1998) and Grimes et al. (2004).

The empirical \oiii\ LF found by S05 is in contradiction to the simple receding torus  model and the
paper proceeds to examine the various possibilities for the discrepancy. In
particular it shows that some of the discrepancy can be explained  by
neglecting the assumption of a luminosity independent \LOLX. Specifically,
a relationship of the type 
\begin{equation}
L({\rm [OIII]}\,\lambda5007) \propto L_{rad}^{1-2 \xi} \,\, ,
\end{equation}
where $L_{rad}$ is the total radiated continuum luminosity that heats the dust in the torus,
 with $\xi \simeq 0.23$, can lead to a much better agreement
between model and observations. The author then dismisses this idea on grounds of the well known near-constancy
of EW(\oiii) in QSOs based on the Miller et al. (1992) results
 and the known proportionalities between L(\oiii) and other quantities based on M94 and Grimes et al (2004).
As shown here, some of those early claims are not substantiated by our new analysis of
X-ray selected type-II and optically selected type-I AGN samples.
In fact, our Fig. 1 and eqns. 1 \& 2  show relationships  of
exactly  the type suggested by S05.
Thus part of the  discrepancy can be explained in this way. This
cannot be the entire explanation since our data also show a shift in the mean
\LOLX\ between the two AGN groups. We suggest that the S05 calculations  be repeated 
 taking into account all those effects, including the possibility of disappearing NLRs at high continuum luminosities.
A full investigation of this type is beyond the scope of the present paper.

\acknowledgements

We are grateful to R. Maiolino, G. Szokoly, 
M. Crenshaw, S. Kraemer, G. Hasinger, J. Turner,
G. Kauffmann and Marcella Brusa for helpful comments and discussions. 
We also benefitted from the useful comments of the referee, M. Strauss. 
Funding for the creation and distribution of the SDSS Archive has been
provided by the Alfred P. Sloan Foundation, the Participating
Institutions, the National Aeronautics and Space Administration, the
National Science Foundation, the U.S. Department of Energy, the
Japanese Monbukagakusho, and the Max Planck Society.
The SDSS Web site is http://www.sdss.org/.
The SDSS is managed by the Astrophysical Research Consortium (ARC) for
the Participating Institutions. The Participating Institutions are The
University of Chicago, Fermilab, the Institute for Advanced Study, the
Japan Participation Group, The Johns Hopkins University, Los Alamos
National Laboratory, the Max-Planck-Institute for Astronomy (MPIA),
the Max-Planck-Institute for Astrophysics (MPA), New Mexico State
University, University of Pittsburgh, Princeton University, the United
States Naval Observatory, and the University of Washington.
This research has made use of the NED database which is operated by
the Jet Propulsion Laboratory, California Institute of Technology,
under contract with the National Aeronautics and Space Administration.
This work is supported by the Israel Science Foundation grant 232/03
and by the chair of Extragalactic Astronomy at Tel Aviv University.
HN acknowledges an Humboldt foundation prize and thank the host institution, MPE Garching, 
where most of this work has been done.

\end{document}